# Plate Acoustic Waves in ZX-cut Lithium Niobate.

I. Ostrovskii, A. Nadtochiy, V. Klymko.

Department of Physics and Astronomy, The University of Mississippi

(iostrov@phy.olemiss.edu)

*Abstract* – **Plate acoustic waves (PAW) propagating along X-axis in the Z-cut wafer of a single crystal of lithium niobate are considered theoretically and experimentally. For eight lowest PAW modes, the dispersion curves for wavenumber *k*(*f*) are calculated by the equations of motion and electrodynamics, by the Finite Element Method, and then measured experimentally. The spectra *k*(*f*) obtained by the numerical solution and FEM-simulation are in good agreement, and experimental measurements agree with theoretical predictions. The PAW modes are identified by the components of their total acoustic displacements and cutoff frequencies. Analysis of the longitudinal and normal acoustical displacements permits to find PAW mode capable for usage in ultrasonic actuators. The results obtained may be useful for ultrasonic transducers, acousto-electronic and acousto-optic applications, and u̲ltrasonic motors/actuators fabricated in the Z-cut ferroelectric lithium niobate wafers including periodically poled wave-guides.**

I.    INTRODUCTION

The plate acoustic waves (PAW) that propagate in thin ferroelectric waveguides are an important part of Ultrasonics and Ferroelectrics. Basic properties of PAW in isotropic media are described in mid 60-ths [1], [2]. Later, PAW was excited in crystalline plates. Among the first publications in international Journals, we mention the excitation of Lamb waves and observation of nonlinear acousto-electric effects in a single crystal plate of CdS [3], [4]. Further, PAW in ferroelectric lithium niobate plates were excited and used for observation of the new effects of acousto-electric injection [5] and surface crystal sonoluminescence [6]. In general, PAWs have diverse applications including electronics, acousto-electronics, acousto-optics, and various engineering devices. A progress in this field is reflected in many recent publications. For instance, Lamb waves in ZY- and YZ-cut of lithium niobate [7], quasi-shear horizontal acoustic wave in ZX-cut lithium niobate [8], laser generation of antisymmetric Lamb waves in isotropic plates [9], and Lamb waves in periodically corrugated isotropic plates were investigated [10], [11]. In Ultrasonics, Plate Waves are used in the delay lines and piezoelectric acoustic transducers [12]-[15]. Among new applications, the zero order PAW modes are employed in the small mass detectors [16]. Besides ferroelectrics, the properties of PAW are investigated in different materials such as Y-cut and ST-cut quartz [17], [18], and potassium niobate [19].

Usually, the YZ-cut crystallographic orientation of lithium niobate is under study, where PAW propagates along the Z-axis in the Y-cut wafer. This traditional choice is due to long-term works on the Interdigital Transducers, acousto-electric devices, and Surface Acoustic Waves propagating in that very cut. The dispersion curves for some modes of PAW in lithium niobate (LN) plate are considered in references [20], [21], but not ZX-cut. As for other crystallographic cuts, the dispersion of phase velocity in LN chip is calculated for Y-cut with Z-propagation [22], and for the X-cut with Y+30º propagation, and the Y-cut with X propagation [23]. However, a phase velocity dispersion of X-propagating PAW in Z-cut wafer was not considered yet.

To characterize piezoelectric properties of PAW, the effective electromechanical coupling coefficient of lithium niobate, lithium tantalite [24], [25], and other piezoelectric materials [26] is investigated. However, in reference [24] the dispersion curves for PAW modes are not calculated for any one crystallographic orientation of LN wafer, and in reference [25] Z-cut wafer are not considered at all.

Different analytical methods were proposed to analyze the propagation of plate waves, such as potential field and normal mode methods [20], [2]. For theoretical calculations, an alternative to an analytical solution is an approach using the Finite Element Method (FEM) for numerical simulation [27], [28] of an acoustic problem. At this time, we are not aware of any published results for LN crystalline plate calculated by using the FEM model.

Surprisingly, another important crystallographic cut of LN, ZX-cut, was not considered for propagation of PAW modes in Z-cut wafer along the X-direction. The Z-cut wafers are extensively used for fabrication of the periodically poled LN or lithium tantalite, for further optical/laser harmonics generation, for nonlinear acousto-optic phenomena, etc. [29]-[31]. The ZX-cut especially is important for fabrication of Ti:LiNbO$_3$ optical waveguides [32]. In addition, ZX-cut may be useful for fabrication of the periodically poled ultrasonic vibrators [33] operating at the "domain resonance" [34]. To the best of our knowledge, we can not refer to any known publication or presentation describing theoretical and experimental results on the dispersion curves of X-propagating PAW in a Z-cut LN wafer.

In this work, we report the results on propagation along X-axis of eight lowest PAW modes in the Z-cut of LN wafer including the theoretical an experimental synthetic spectra *k(f)*. The particular goals are to consider PAW in the same crystallographic ZX-cut of LiNbO$_3$ that is used for nonlinear optical harmonic generation in periodically poled waveguides, to verify applicability of the FEM simulation to waves in crystalline plates, to analyze acoustic modes identification by their displacements, and to find PAW-mode for applications in nanotechnology for nano/micro particles transport.

## II. ANALYTICAL-NUMERICAL DISPERSION CURVES

We calculated the analytical-numerical dispersion curves by using representation of the sought solution as the sum of the partial waves [7], [22], [25]. This is done to provide verification for the FEM model. The equations of motion for piezoelectric plate along with the equations of electrodynamics and corresponding boundary conditions were solved for our crystallographic orientation shown in fig. 1.

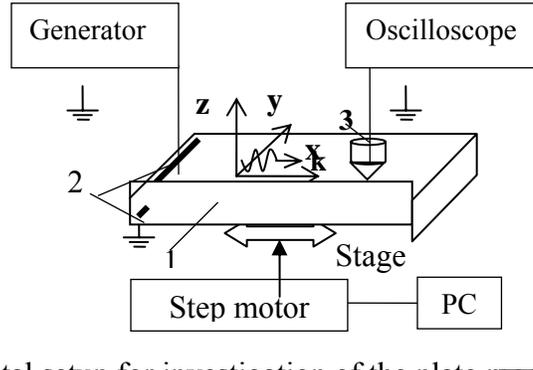

**Figure 1.** Experimental setup for investigation of the plate modes in the Z-cut LN wafer.

*1* – lithium niobate wafer, *2* – input transducer, *3* – output pickup.

The analytical solutions are calculated by the equations of motion (1) and electrodynamics (2) that describe PAW propagation in a ferroelectric:

$$\rho \frac{\partial^2 u_i}{\partial t^2} = \frac{\partial T_{ij}}{\partial x_j} = \frac{\partial}{\partial x_j}(c_{ijkl}^E \frac{\partial u_k}{\partial x_l} - e_{m\,ij} E_m) \qquad (1)$$

$$\frac{\partial D_i}{\partial x_i} = \frac{\partial}{\partial x_i}(e_{ikl}\frac{\partial u_k}{\partial x_l} + \varepsilon_{ij}^S E_j) = 0 \, , \qquad (2)$$

where $u_i$ are the components of acoustic displacement, $T_{ij}$ is mechanical stress tensor, $E_i$ and $D_i$ are electric field and displacement, respectively, $c_{ijkl}^E$ is elastic modules tensor, $e_{ikl}$ and $\varepsilon_{ij}^S$ are piezoelectric and dielectric constant tensors, respectively, and indexes $i, j, k, l$ are running from 1 to 3. Electric field depends on electric potential $\varphi$ as $\mathbf{E} = -\nabla \varphi$. At the plate surfaces $z = \pm h$, the boundary conditions of zero acoustic stress (3) and continuity of normal component of electrical displacement (4) are applied.

$$T_{iz}\big|_{z=\pm h} = 0, \qquad i = x, y, z \tag{3}$$

$$D_{z,crystal}\big|_{z=\pm h} = D_{z,air}\big|_{z=\pm h}. \tag{4}$$

The components of the mechanical stress tensor $T_{ij}$ and electric displacement $D_i$ are defined in the parenthesis of right parts of the equations (1) and (2), respectively. The lithium niobate crystal is an anisotropic media, in which each propagating PAW has all the three components of mechanical displacement. For instance, the waves propagating along the $x$-axis have longitudinal $u_x$, shear horizontal $u_y$, and shear normal $u_z$ components. For this case, the solutions of the equations (1) through (4) for the displacement and potential may be written in a general form:

$$u_i = u_{i0} \exp i(\omega t - 2\pi k x - 2\pi k \beta z), \tag{5}$$

where wave number $k=(1/\lambda)$ is inverse wavelength of PAW, and parameter β gives distribution of the amplitudes $u_{i0}$ along Z-axis. The index $i = 1, 2$ and 3 for the acoustic displacements along $x, y$ and $z$ axes, respectively, and $i = 4$ for electric potential $\varphi$. The choice of the $k$ being inverse wavelength is useful for further computations and following comparison of the results obtained by analytical solution and Finite Element Method simulation. The general solutions (5) are substituted into (1) and (2) in view of (3) and (4). The determinant of the resulting system of equations is a polynomial of eights degree in $\beta$. As a result, the parameter $\beta$ is obtained as a function of a wave number $k$. Next step is to apply the boundary conditions (3) and (4) by using the acoustic solutions (5) with known $\beta$, and electric field $E$, which in turn is calculated by equation (2) in terms of acoustic displacement. Three equations (3) yield six boundary conditions and equation (4) yields two additional equalities. Last set of eight boundary condition equations is numerically solved, and the dependencies of inverse wavelength *k* versus frequency *f* for several PAW modes are obtained. The magnitudes of the elastic, piezoelectric, dielectric constants and material density that we use in our calculations are taken from well-known monograph [20]. The computed dispersion curves for lowest eight plate modes propagating along the X- axis are presented in fig. 2 by the solid black

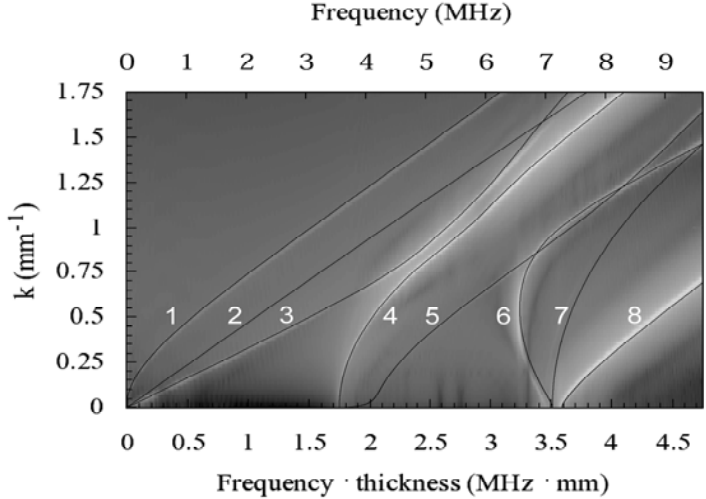

lines.

**Figure 2.** Analytical dispersion curves (thin black lines) and synthetic dispersion spectrum (white lines, details in section V) of PAW propagating along *x*-axis in the Z-cut LN wafer.

III. THE FEM MODEL

In this work, we want to investigate both propagation and excitation of the PAW in Z-cut lithium niobate wafer. This double problem with electro-acoustic transformation at the location of a single metal electrode transducer may be solved by using the well-known Finite Element Method (FEM) [35]-[38]. In the case of PAW, we use the two-dimensional (2-D) numerical FEM-model with the coordinate system specified in fig. 1. The plate is considered infinite in the *y* direction for PAW propagating along the *x*-axis. The 13x1501 nodes non-uniform mesh is used to represent a wafer cross-section in X-Z plane, and an air above and below the wafer. The mesh consists of linear triangle elements. In the crystal, elements have equal sides of length $h_m = 0.05$ mm along the *x* and *z* directions. The mesh is extended into the air by the distance equal to ten plate thicknesses. The size of elements in the *z* direction in the air is 10 times larger than in lithium niobate. The element size in the *x* direction is the same in the air and in the crystal. This non-uniform mesh is introduced to include into the calculations the electric field in the air, while saving the computational resources. Our simulations shown that 10·(2h)-thick air surrounding, where (2h) is the thickness of the plate, is actually enough for an accurate solution; further increase in the air thickness did not change the spectra of plate waves in a wafer. As an alternative method, the representation of exterior region as a single "super-element" may be used [38]. However, the mesh extension algorithm is applied in this work because it does not require any additional computations. The energy of electric field in the air decreases exponentially as $\sim e^{-2k \cdot d}$. At the distance $d = 10 \cdot 0.5$ mm the wave with the non-angular wavenumber $k = 0.5$ mm$^{-1}$ carries less than 1% of its initial energy on the crystal surface. Thus by choosing 10·(2h)-thick air slabs, we make a 1% error, which is quite acceptable for FEM simulations. Further, to model PAW excitation in FEM-computations, we assign certain values of an electric potential at crystal surface at the location of metal electrode (2 in fig. 1). In the calculations, the rf-voltages of 1, 10, 10$^2$, and 10$^3$ Volts were used. The synthetic dispersion curves obtained under different levels of excitation were identical.

The energy loss in the plate is modeled by introducing the imaginary part to the elastic constants $c_{ijkl} = c'_{ijkl} + i c''_{ijkl}$. We take the imaginary part $c''$ equals to 0.1% of the real part that corresponds to typical mechanical quality of a LN-crystal of the order of 10$^3$. Then $c'' = \delta\, c'$, where parameter δ = 0.001. To eliminate the reflections from the waveguide edges under FEM calculations, we add in the model the absorbing loads to both ends of the plate. These loads consist of a number $n$ of absorbing layers. The magnitude of the imaginary part in $n$-th load layer depends on $n$ as $c''_n = \delta\, c'(\gamma)^n$, where $\gamma$ is a numeric parameter. Total number of layers is 100 ($n$ = 1, 2, ….100.). Computations show that with the parameter γ = 1.07 we can practically minimize to zero the reflections of PAW from the crystal end faces.

The numerical model calculates the amplitude of acoustical displacements $u_x(f_j, x_m)$, $u_y(f_j, x_m)$, and $u_z(f_j, x_m)$ for the set of frequencies $f_j$ at nodes $x_m$ on the plate surface along the direction of propagation. The discrete spatial Fourier transform is applied to obtain the Fourier image of the displacements $\hat{u}_{x,y,z}(f_j, k_i)$ in the frequency-wavenumber domain. The synthetic dispersion spectrum is the grayscale plot of the Fourier images of total acoustical displacements $\hat{u}(f_j, k_i)$. The total acoustic displacement is defined by following equation.

$$\hat{u}(f_j, k_i) = \sqrt{\left[\hat{u}_x(f_j, k_i)\right]^2 + \left[\hat{u}_y(f_j, k_i)\right]^2 + \left[\hat{u}_z(f_j, k_i)\right]^2}. \qquad (6)$$

To reduce approximation error, it is required [37] that $\lambda/h_m > 10$, where λ is acoustic wavelength and $h_m$ is the element size inside the plate. This requirement is satisfied in our model for $k < 2$ mm$^{-1}$. The synthetic dispersion spectra calculated by FEM is shown in fig. 2 for the wave number in the range of $0 < k < 1.75$ mm$^{-1}$.

IV. EXPERIMENTAL DETAILS

The experimental setup for excitation and detection of PAW in LN Z-cut plate is shown in fig. 1. The initial wafer was purchased in the market place from MTI Corporation. It is a high quality optical grade LiNbO$_3$ substrate of 3-inch diameter and 0.5 mm thickness. The accuracy of crystallographic cut is <5′ for z-axis, and <10′′ for x- and y- axes. The accuracy of plate thickness is about 1%. In view of the fact that plate thickness is very important for further analysis of the acoustic mode spectra and mode identification, we measure the thickness by the frequencies of wafer resonance vibrations along z-axis. It yields the exact plate thickness of 0.493 mm. Further, we take the experimental data from the 40-mm long central part of the wafer, both surfaces of which are optically polished. This part of the wafer is shown in fig. 1 as plate 1. In all measurements, the wafer is suspended in the air. The end edges of the plate are placed on the supports, which are fixed to a moving table. In fig. 1, the crystallographic z-axis is normal to wafer surface. In our experiments, the rf-voltage applied to the input single-electrode transducer *2* excites acoustic waves in the plate. Two types of the input voltage used in the experiments are the 1 kHz meander and burst filled with rf-sine frequencies from 3.5 MHz to 8 MHz. The input transducer consists of two 0.5 mm wide and 2 cm long metal electrodes deposited on the plate surfaces one across another. The bottom electrode in fig. 1 is grounded. The PAW propagates along crystallographic x-axis. In order to obtain the experimental synthetic dispersion spectra, the time-dependent acoustical displacement $u_z(t_n, x_m)$ and electric potential φ($t_n, x_m$) are measured at a set of discrete points $x_m$ onto plate surface. The acoustical displacement is measured when PAW is excited by the sine burst input signal. The electric potential is measured when the waves are excited by the meander input signal. The choices of the excitation signals are because of different output sensors. The type of the output pick up depends on the type of detected signal. The acoustic displacement $u_z(t_n, x_m)$ is picked up by a piezoelectric transducer, and electric potential φ($t_n$, $x_m$) is detected by a narrow metal electrode. In the experiment, the output detector *3* is set on the sample surface. The sample itself rests on the stage and it is translated along the *x*-axis. The moving table is powered by a PC controlled step motor. The discrete readings of the detector 3 taken at *N* different times $t_n$ (*n* = 1, 2,…N) are recorded at each stop point $x_m$ (*m* = 1, 2… *M*). Then, the discrete double Fourier transform is applied to the *N×M* matrix of the output data. The result of the double Fourier transform is the discrete function û($f_j$, $k_i$) depending on the frequency and wave number. The magnitudes of û($f_j$, $k_i$) function represent the amplitudes of the corresponding acoustic modes. A gray-scale coded plot of the function û($f_j$, $k_i$) in frequency — wavenumber domain gives a set of experimental synthetic dispersion curves.

## V.     RESULTS AND DISCUSSION

The synthetic dispersion spectra of PAW propagating along the *x*-axis in Z-cut LN wafer are presented in fig. 2 in a gray scale. The curves in figure 2 are numbered from 1 through 8 to denote different propagating PAW modes. The white and light gray curves on the dark gray background depict the synthetic dispersion spectrum *k*(*f*) obtained from the numerical FEM simulation. The intensity of white color is proportional to the amplitude of Fourier image û($f_j$, $k_i$) of corresponding acoustical displacement. The superimposed black curves represent the analytical solution. There is good agreement between the two families of dispersion curves in fig. 2, which proofs our FEM calculations.

For PAW modes propagating along the *x*-axis, we have the longitudinal $u_x$, shear normal $u_z$, and shear horizontal $u_y$ displacements. In the analytical solution, the order of modes is identified by the number of half wavelengths along the plate thickness [1], [2]. A quasi-Lamb mode has two main components of displacement along the *x*- and *z*-axes, which are accompanied by a relatively small shear component along the *y*-axis. We denote the quasi-Lamb modes having symmetrical displacement $u_z$ with respect to sagittal plane (*z* = 0) by *"S"*, and call them symmetrical Lamb waves. Those modes with antisymmetrical distribution of $u_z$ with respect to sagittal plane are denoted by *"A"* and called antisymmetrical Lamb waves. Mathematically, the S-modes have $u_z$ ~ Sin(*k* ·*β* ·*z*) and A-modes have their $u_z$ ~ Cos(*k* ·*β* ·*z*), where parameter *β* depends on modes' frequency and phase speed [39]. The result is, in the S-modes, the *z*- displacements $u_z$ are directed to the two plate surfaces out of the sagittal plane at $z = 0$ giving stretching or compression a waveguide along $z$-axis. In contrary, in the case of $A$-modes, the waveguide is bending up and down progressively as mode propagates along the plate.

Quasi-shear horizontal (*SH*) modes propagating along the $x$-axis have the main displacement component in the $y$ crystallographic direction. The *SH* modes with symmetrical distribution of $u_y$ along $z$-axis are called shear symmetrical (*SS*) modes, and those having antisymmetrical distribution of $u_y$ along $z$-axis are called shear antisymmetrical (*SA*) modes. Mathematically, *SS*-modes have $u_y \sim \mathrm{Cos}(k \cdot \alpha \cdot z)$, and *SA* modes have $u_y \sim \mathrm{Sin}(k \cdot \alpha \cdot z)$, where parameter $\alpha$ is a function of frequency and phase speed [39]. The three zero modes $A_0$, $S_0$ and $SS_0$ occur at zero frequency, they may propagate in any direction in $z$-plane. The higher order modes at their cutoff frequencies are standing waves along the $z$-direction, which implies zero group velocities along the $x$-axis. We see this result in fig. 2 for PAW propagating along $x$-axis, where at the cutoff frequencies the modes 4 through 8 have the wave numbers close to $k = 0$, and corresponding analysis shows the group velocities ($d\omega/dk$) = 0. With the frequency increase higher then cutoff, the wave number of a given mode $k$ increases as well.

To identify properly the PAW modes, one can analyze the acoustic displacements on wafer surface for each mode near its cutoff frequency. We calculated analytically the amplitudes along all three directions near cutoff frequencies when wave number $k$ is very small, $k = 0.1$ (Table 1). We note, this method is unlike to another one proposed in [1], [2] for isotropic plates; in the later it is taken into account a number of bulk acoustic wave half-wavelengths along plate thickness. However, our comparison shows a good agreement between the two approaches.

**Table 1.** Analytical identification of X-propagating PAW modes in the Z-cut LN at $k = 0.1$.

| Mode number | f (MHz) | k (mm$^{-1}$) | $u_x$ | $u_y$ | $u_z$ | Mode type |
|---|---|---|---|---|---|---|
| 1 | 0.0560 | 0.10 | 0.155 | 0.001 | 1 | $A_0$ |
| 2 | 0.4376 | 0.10 | 0.020 | 1 | 0.001 | $SS_0$ |
| 3 | 0.6268 | 0.10 | 1 | 0.024 | 0.048 | $S_0$ |
| 4 | 3.6453 | 0.10 | 0.023 | 1 | 0.001 | $SA_1$ |
| 5 | 4.3082 | 0.10 | 1 | 0.010 | 0.098 | $A_1$ |

| 6 | 7.1022 | 0.10 | 1 | 0.023 | 0.720 | $S_1$ |
| 7 | 7.2573 | 0.10 | 0.023 | 1 | 0.001 | $SS_1$ |
| 8 | 7.6129 | 0.10 | 0.699 | 0.017 | 1 | $S_2$ |

The mode identification in Table 1 is well supported by good correlation with the FEM simulations presented in Table 2. The data of Table 2 are computed for the frequencies close to modes occurrence with the same small wave number k = 0.10 as in Table 1. Comparison of the data in Tables 1 and 2 reveal good agreement between analytical calculations and computer simulation by the FEM.

**Table 2.** FEM identification of *x*-propagating PAW modes in the Z-cut LN at *k* =0.1.

| Mode number | f (MHz) | $k$ (mm$^{-1}$) | $u_x$ | $u_y$ | $u_z$ | Mode type |
|---|---|---|---|---|---|---|
| 1 | 0.06 | 0.1 | 0.146 | 0.001 | 1 | $A_0$ |
| 2 | 0.44 | 0.1 | 0.025 | 1 | 0.001 | $SS_0$ |
| 3 | 0.63 | 0.1 | 1 | 0.02 | 0.049 | $S_0$ |
| 4 | 3.65 | 0.1 | 0.023 | 1 | 0.003 | $SA_1$ |
| 5 | 4.31 | 0.1 | 1 | 0.01 | 0.101 | $A_1$ |
| 6 | 7.10 | 0.1 | 1 | 0.03 | 0.697 | $S_1$ |
| 7 | 7.26 | 0.1 | 0.058 | 1 | 0.025 | $SS_1$ |
| 8 | 7.61 | 0.1 | 0.541 | 0.013 | 1 | $S_2$ |

In the Tables 1 and 2, the maximum amplitude component along the *x*- or *y*-, or *z*-direction for a given mode is taken exactly one (1.000), and then other displacement components along two remaining directions are the fractions of the dominant component. As frequency is increasing above the cutoff one, the acoustic displacement in a corresponding mode is composed mainly of three components $u_x$, $u_y$ and $u_z$. However, a dominant displacement with the highest amplitude remains along the same crystallographic direction as in Tables 1, 2. Unlike other modes, the first symmetrical Lamb mode $S_1$ has a negative group velocity ($d\omega/dk$) near its cutoff in the frequency range from 7.24 MHz down to 6.6 MHz.

Overall, there is good agreement in fig. 2 between the FEM simulation (gray scale dispersion curves) and analytically calculated dispersion curves (solid black lines). In fig. 2, the first order shear antisymmetric mode $SA_1$ (curve 4) is clearly observable in the FEM solution. The first antisymmetric Lamb mode $A_1$ (curve 5) arises at 3.62 MHz. The FEM spectrum for this mode is noticeable and consistent with analytics. The first order symmetric Lamb wave $S_1$ (curve 6) and shear mode $SS_1$ (curve 7) have the same cutoff frequency. The analytical value for the cutoff frequency is 7.24 MHz. The dispersion curve of $S_1$ mode has two branches, with positive and negative group velocities, which is similar to the case described in references [40], [41]. The mode $S_2$ (curve 8) has the cutoff near 7.42 MHz, and it is efficiently excited by one electrode transducer. There are several points in fig. 2, where the dispersion curves cross each other. At these points, two modes have the same wave numbers and frequencies. For instance, the dispersion curves of $A_1$ and $S_1$ modes (curves 5 and 6, respectively) cross each other at frequencies 6.9 MHz and 8.42 MHz, and the plots of $S_1$ and $SS_1$ modes (curves 6 and 7, respectively) overlap at 9.75 MHz. The experimental synthetic dispersion spectra of PAW propagating along the *x*-axis in the Z-cut LN wafer are shown in figures 3 and 4.

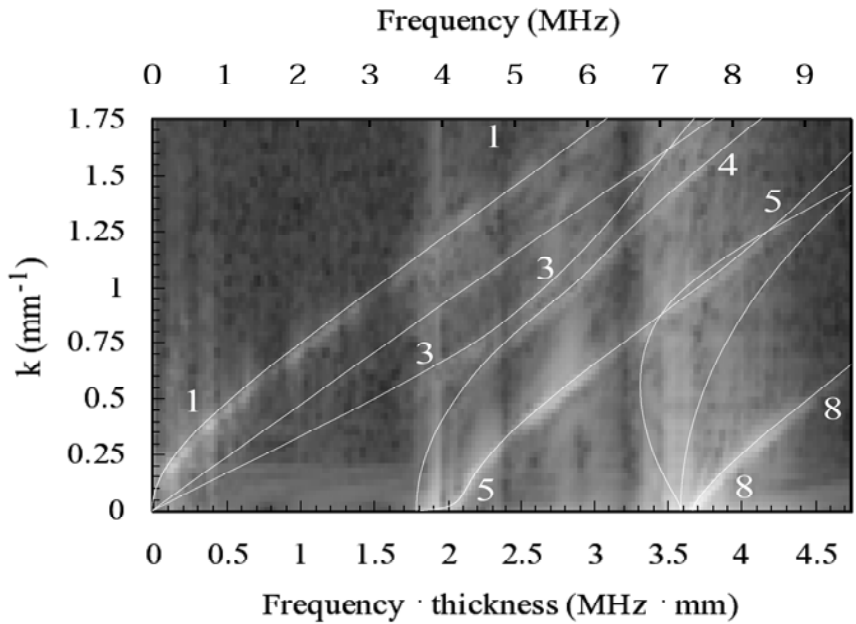

**Figure 3.** Experimental synthetic dispersion spectrum of PAW propagating along *x*-axis in the Z-cut LN wafer (white lines) obtained from the measurements of acoustical displacement. The white hairlines show the analytical dispersion curves, as a guide for eye.

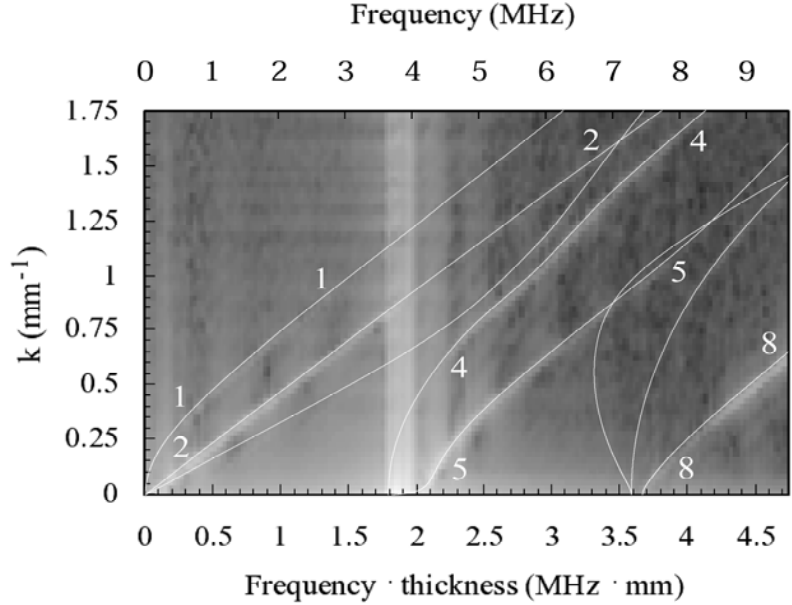

**Figure 4.** Experimental synthetic dispersion spectrum of PAW propagating along *x*-axis in the Z-cut LN wafer (white lines) obtained from the measurements of surface electric potential. The white hairlines show the analytical dispersion curves, as a guide for eye.

The superimposed white hairlines represent the analytical solution. The spectrum of fig. 3 is obtained by measuring the acoustic displacements onto sample surface and those in fig. 4 – by detection of the surface electric potential. In figures 3 and 4, the experimental dispersion plots that were reliably detected are shown in between corresponding numbers indicating the PAW mode. For example, in figs. 3 and 4, second symmetrical mode $S_2$ was detected at those frequencies and wave numbers that are in between two numbers 8 (since $S_2$ mode is numbered 8), etc. The absence of the mode number in figs. 3, 4 means the corresponding PAW mode is not reliably detected in our experiments.

The zero symmetrical mode $A_0$ (plot 1) is effectively excited in a frequency range from zero to 6 MHz. The zero shear horizontal mode $SS_0$ (plot 2) is efficiently detected from 4 MHz to 6.5 MHz. The zero symmetric mode $S_0$ (plot 3) is inefficiently excited only at the frequencies from 4 MHz to 5.5 MHz. The other modes detected by their acoustic displacements reveal themselves in the following frequency ranges: $SA_1$ (plot 4) from 4.5 to 8 MHz, $A_1$ (plot 5) from 3.75 to 9 MHz, $S_2$ (plot 8) - 7.3 to 9.5 MHz. The modes $S_1$ (plot 6 in fig. 2) and $SS_1$ (plot 7 in fig. 2) are not reliably detected, and they are absent in fig. 3. It means the normal component $u_z$ in these modes does not appear strong enough at crystal surface, which may be connected to a specific character of the displacement distribution over plate thickness. The first order antisymmetrical mode $A_1$ (plot 5) gets the most effective excitation by the single electrode transducer.

In addition to the propagating PAW modes denoted by numbers 1 through 5 in figs. 3 and 4, one can also notice the traces of vertical lines near 3.6 - 4 and 7.3 MHz. This is a reaction of the pick-up on the standing vibrations of the wafer. Indeed, the speed of two transverse bulk waves ($V_T$) propagating along z-direction and having the displacements along x- or y-axes is 3.574 km/s [42]. The Z-cut LN wafer used in our experiments has thickness of 0.493 mm. It exactly corresponds to the half-wavelength resonance for the transverse acoustic waves at frequency of 3.623 MHz. In the experiments, the pick up detects this thickness resonance right near 3.6 MHz (figures 3, 4). Similarly, at 7.3 MHz, the sample thickness is exactly equal to a half wavelength of bulk longitudinal wave ($V_L$) propagating along z-axis with the speed of 7.3 km/s [43]. This longitudinal wave resonance is picked up at 7.3 MHz in fig. 3.

As mentioned above, the second series of the experimental measurements involves detection of the electric potential onto the sample surface. The corresponding experimental synthetic dispersion spectrum is shown in fig. 4. Three zero order modes and four higher order modes were reliably detected by there electric potential onto sample surface. The zero antisymmetric mode $A_0$ (plot 1) is revealed at frequencies from 0.75 to 3.0 MHz. The zero shear horizontal symmetrical mode $SS_0$ (plot 2) is detected almost over all frequency range from 0 to 7 MHz. The zero symmetrical $S_0$ mode is read only within a narrow frequency range near 6 MHz (plot 3), although its amplitude is weak. The first order modes (plots 4 and 5) have the same cutoff frequencies near 3.6 MHz. Lamb mode $S_1$ numbered 6 in figs. 2, 3 was not detected in this experiment and does not appear in fig. 4. Shear mode $SS_1$ numbered 7 becomes detectable at the frequency of about 8 MHz. At last, Lamb mode $S_2$ (plot 8) is detected at all frequencies from its cutoff near 7.4 MHz up to 10 MHz.

*In summary*: Theoretical and experimental investigations of the plate acoustic waves propagating along x-axis in the Z-cut Lithium Niobate suggest the following conclusions.

1) Analytical-numerical solutions, Finite Element Method simulation, and experimental data on the dispersion curves for the first eight modes are in good agreement. The PAW modes can be identified by their x-, y-, z- displacement components near modes' cutoff frequency. It is particularly useful for identification of PAW modes that have the same cutoff frequencies, in which case a traditional method of counting the number of wavelengths along plate thickness may lead to confusion.

2) Identification of PAW-modes by their amplitude components along $x$-, $y$-, and $z$ – axes permits to pick up the $S_1$ and $S_2$ modes as the most promising for applications as an ultrasonic transporter of micro/nano-particles because these modes have respectively large amplitudes of both normal $u_z$ and longitudinal $u_x$ amplitudes near cutoff frequencies. Numerical modeling and experimental investigation of the eight lowest modes suggest that mode $S_2$ is most effectively excited. Thus, $S_2$ mode can be used in the applications such as ultrasonic transporter of micro/nano-particles or ultrasonic actuator.

3) The Finite Element Method simulation is shown to be applicable for solving the acousto-electric problem on propagation of dispersive Plate Waves in ferroelectric wafers. The synthetic dispersion spectrum $k(f)$ obtained from the FEM simulation provides information about the dispersion of PAW modes and amplitudes of displacement components. The dispersion curves obtained by the FEM and analytical considerations turn to be mostly identical. However, a small discrepancy of 5% or less is observed at those frequencies and wavenumber, where more than one acoustic mode do exist, for instance, near cutoff frequency of 7.1 MHz (3.5 MHz·mm).

4) At low frequency limit, when acoustic wavelength is much longer than a plate thickness, three zero modes $A_0$, $SS_0$, and $S_0$ can be excited. The zero symmetrical mode $SS_0$ is non-dispersive shear horizontal wave. These modes have a dominant acoustic displacement.

5) Besides three zero modes, as frequency increase other modes occur in the following order: $SA_1$, $A_1$, $S_1$, $SS_1$, and $S_2$. These modes have their cutoff frequencies when a wafer thickness is equal to an integral number of half wavelengths of corresponding bulk acoustic wave. For instance, the antisymmetrical modes $SA_1$ and $A_1$ occur at the frequency $f = f_T$, when $(2h) = 0.5 \cdot \lambda_T = 0.5 \ V_T/f_T$, the first symmetrical modes $S_1$ and $SS_1$ emerge at frequency $f = 2 \cdot f_T$ when $(2h) = \lambda_T$, and the second symmetrical mode $S_2$ shows up at frequency $f = f_L$, when $(2h) = 0.5 \cdot \lambda_L = 0.5 \ V_L/f_L$.

6) The results described in this work may be useful for developing various acousto-optic and acousto-electronic devices, and Ultrasonic Motors or Actuators that are designed on the Z-cut Lithium Niobate wafers including periodically poled crystals.